\documentclass[prd, preprint, 12pt]{revtex4-1}

\usepackage{amsmath}
\usepackage{amssymb}
\usepackage{setspace}
\usepackage{graphicx}
\usepackage{natbib}
\usepackage{float}

\begin{document}
 
 %

\begin{center}
 { \large {\bf A new length scale for quantum gravity\\ \it - and a resolution of the black hole information loss paradox - }}


\vskip 0.3 in

{\large{\bf Tejinder P.  Singh}}

{\it Tata Institute of Fundamental Research,}
{\it Homi Bhabha Road, Mumbai 400005, India}\\
\bigskip
{\tt tpsingh@tifr.res.in}\\

\end{center}

\bigskip
\bigskip

\centerline{\bf ABSTRACT}

\bigskip
\noindent We show why and how Compton wavelength and Schwarzschild radius should be combined into one single new length scale, which we call the Compton-Schwarzschild length.  Doing so offers a resolution of the black hole information loss paradox, and suggests Planck mass remnant black holes as candidates for dark matter. It also compels us to introduce torsion, and identify the Dirac field with a complex torsion field. Dirac equation, and Einstein equations, are shown to be mutually dual limiting cases of an underlying gravitation theory which involves the Compton-Schwarzschild length scale, and includes a complex torsion field.

\noindent 

\vskip 1 in

\centerline{March 31, 2017}

\bigskip

\centerline{Essay written for the Gravity Research Foundation 2017 Awards for Essays on Gravitation}
\bigskip

\bigskip
\centerline{\it This essay received an honorable mention in the Gravity Research Foundation 2017 Essay Contest}

\newpage

\setstretch{1.3}

\bigskip

\noindent A relativistic particle with a given mass $m$ has two length scales associated with it: the half-Compton wavelength $\lambda_{C}=\hbar/2mc$, and the Schwarzschild radius $R_S=2Gm/c^2$. It however does not `know' whether it should obey the Dirac equation of relativistic quantum theory, or the Einstein equations of general relativity. Neither of the two theories give any indication to this effect, and both claim to hold for all values of $m$.  We know only from experiments that the Dirac equation holds if $m\ll m_{Pl}$, i.e. if $\lambda_C \gg L_{Pl}$, where $m_{Pl}$ and $L_{Pl}$ are respectively the Planck mass and Planck length.  Einstein equations hold if $m\gg m_{Pl}$ i.e. $R_S\gg L_{Pl}$.  From the theoretical viewpoint, it is unsatisfactory that the two theories should have to depend on experiment in order to establish their domain of validity. There ought to exist a more general description, valid for all $m$, and in particular when $m\sim m_{Pl}$, and  $R_S\sim \lambda_{C}\sim L_{Pl}$; with the Dirac equation and  Einstein equations emerging as limiting cases. The need for such a description is also necessitated if we make the plausible assumption that Planck length is the smallest physically meaningful length. It is then physically unreasonable to talk of a $R_S < L_{Pl}$ 
when $m<m_{Pl}$, and to talk of a $\lambda_{C}<L_{Pl}$ when $m>m_{Pl}$. It appears more reasonable to have only one universal length associated with a mass $m$, so that this length always stays higher than $L_{Pl}$, irrespective of whether $m$ is greater or smaller than Planck mass. This length should reduce to $R_S$ for $m\gg m_{Pl}$, and to $\lambda_C$ for $m\ll m_{Pl}$ \cite{Singh2008,Singh2014a,Singh2014b,Singh2015,Carr1,Carr2,Carr3}.

One possible way to define such a universal length, which we call the Compton-Schwarzschild length, and which we label $L_{CS}$, is to simply add the Compton wavelength to the Schwarzschild radius:
\begin{equation}
\frac{L_{CS}}{2L_{Pl}} \equiv \frac{1}{2} \left(\frac{2m}{m_{Pl}} + \frac{m_{Pl}}{2m}\right) = \cosh z; \qquad\qquad z\equiv \ln (2m/m_{Pl})
\label{unilength}
\end{equation}
This length is plotted in Fig.1 as a function of the logarithmic mass, and it has interesting properties. It takes the minimum value $2L_{Pl}$ at $m=m_{Pl}/2$, matches with $R_S$ for $m\gg m_{Pl}$, and with $\lambda_C$ for $m\ll m_{Pl}$. For any given value of $L_{CS}$, two values of $m$, say $m_q$ and $m_{c}$, are possible, and they are related as $m_qm_c=m_{Pl}^2/4$. In terms of $z$, we note that there is a reflection symmetry $z\leftrightarrow -z$ about $z=0$. This reflection thus maps the gravity dominated regime to the quantum dominated regime, and vice versa.

There are of course other possible interpolating functions which could define the universal length, besides $\cosh(z)$, and which limit to $R_S$ for large $m$, and to $\lambda_C$ for small $m$. They have a minimum value around $L_{Pl}$, and every value of $L_{CS}$ admits two solutions for $m$, even though they may not have a  $z$ reflection symmetry. Such a universal length has a far-reaching implication for black hole evaporation, and for the information loss paradox. We illustrate this using the form (\ref{unilength}) for the universal length, although the conclusion holds  for a generic interpolating form. 
The dynamical process must now involve, not the length scale $R_S$, but rather the universal length $L_{CS}$. 
An evaporating black hole, which rolls down the right half of the curve ($z>0$) by losing mass via Hawking radiation, must settle down at the minimum length $2L_{Pl}$, which is at $m=m_{Pl}/2$. It cannot rise up to the left side of the curve and emit radiation, which would require $R_S > \lambda_C$. In principle, such a Planck mass remnant could hold all the information  of the initial black hole, thus resolving the paradox. Planck mass remnant black holes have often been suggested as a resolution, but not incontrovertibly so, in the absence of a quantum theory of gravity \cite{Preskill,kinjalk}. Here, we see that the existence of such a universal length inescapably implies that evaporation must terminate at Planck mass, independent of the details of the underlying quantum gravity theory. It is significant that such Planck mass remnants of evaporating primordial black holes have also been suggested as
candidates for dark matter \cite{macgibbon,carr4}.

\bigskip

\begin{figure} [ht]
 { \centerline{\includegraphics{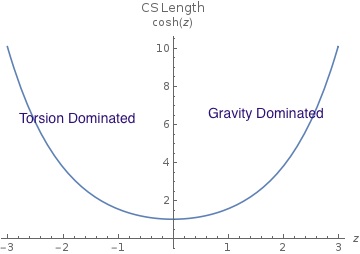}}}
\caption{ Plot of the scaled Compton-Schwarzschild Length $L_{CS}/2L_{Pl}=\cosh (z)$ as a function of 
the logarithmic mass $z=\ln(2m/m_{Pl})$. The CS length $L_{CS}$ attains the minimum value $2L_{Pl}$ 
at $z=0$, i.e. at $m=m_{Pl}/2$. For $z>0$, i.e. $m>m_{Pl}/2$, $L_{CS}$ increases with increasing mass: this is the gravity dominated regime.   For $z<0$, i.e. $m<m_{Pl}/2$, $L_{CS}$ increases with decreasing mass: this is the torsion dominated regime. For any given value of $L_{CS}$, two values of $m$, say $m_q$ and $m_{c}$, are possible, and they are related as $m_qm_c=m_{Pl}^2/4$. [If $\lambda_C$ and $R_S$
were plotted on the same graph, the former comes down along the top left curve but goes to zero as $z\rightarrow\infty$, whereas the latter comes down along the top right curve but goes to zero as $z\rightarrow -\infty$. The universal length superposes these two curves of $\lambda_C$ and $R_S$ and hence invalidates lengths smaller than $2L_{Pl}$ for all $z$].
}
\end{figure}%


The universal length curve helps us search for the underlying theory: the right hand side is the gravity dominated region where gravity is sourced by mass, which enters the Einstein field equations in terms of the length $R_S$. The left hand side of the curve is the quantum dominated region where the Dirac field is also sourced by mass, but it enters the Dirac equation in terms of the length $\lambda_C$. The underlying theory will also be sourced by mass, but the mass will now enter the dynamical equations in terms of the universal length $L_{CS}$. The space-time symmetry group on the left hand is the Poincar\'e group; elementary particles are irreducible representations of the Poincar\'e group, labelled by mass and spin. However on the right hand side,  and this is the puzzle, the local symmetry group [acting on orthonormal frames in the tangent spaces] for the gravity dominated region is {\it not} the Poincar\'e group, but the Lorentz group. Translations are not included. How can it be, that as one moves along the universal length curve, and crosses the minimum at $(m_{Pl},L_{Pl})$, the local symmetry group suddenly changes from Poincar\'e to Lorentz? Or that the dual masses $m_q$ and $m_c$ which both have the same universal length, have different symmetry groups? This appears extremely unnatural and suspicious. A resolution is to generalize the space-time to a Riemann-Cartan space-time, and include torsion, which allows for translations to be included, and converts the local symmetry group on the right hand side to the Poincar\'e group \cite{Trautmann}. But torsion is not observed in the classical world, it vanishes outside matter sources, and is sourced by spin angular momentum \cite{Hehl}. This strongly suggests that torsion is significant on the left hand side, and we identify the Dirac field with a complex torsion field \cite{Singh2014a}! On the right hand side, the matter wave function (represented by the Dirac field) gets extremely localised (classical limit) and hence torsion naturally stays inside the matter source. On the left hand side, gravity is negligible because $m\ll m_{Pl}$, which effectively implies $G\rightarrow 0$. In this way, we can understand the universal length curve and the underlying dynamics in a harmonious symmetric manner.

Let us now heuristically construct an action principle for the underlying theory, which involves the length scale $L_{CS}$, but not $R_S$ and $\lambda_C$, and from which the duality of the Dirac equation  and Einstein equations is manifest. We envisage such an action to be a sum of the Einstein-Hilbert action, the Dirac field action, and a mass-dependent source term which we have to suitably convert to $L_{CS}$. To start with, consider the action as a sum of three terms $T_1, T_2 $ and $T_3$:
\begin{equation}
S = \frac{c^3}{G} \int d^4x \;\sqrt{-g} R + \hbar\int d^{4}x \; \sqrt{-g}\;\overline{\psi}\; i\gamma^{\mu}\partial_\mu\psi
- mc \int d^{4}x\; \sqrt{-g}\; \overline{\psi}{\psi}
\label{schematicaction}
\end{equation}
where the symbols have their usual meaning [strictly speaking, in the second term the gamma matrices and the derivative should be replaced by their curved space counterparts, but that is not of consequence here]. If the first term $T_1$ is absent (flat space-time limit), then the variation of the second and third terms with respect to $\bar{\psi}$ gives the Dirac equation. If the second term $T_2$ is absent (classical limit), and the probability density $\bar{\psi}\psi$ in the third term $T_3$  is replaced by a delta function
$\delta^{3}({\bf x})$, we get upon variation with respect to the metric, Einstein equations for a point mass. If all three terms are present, assume that there is a mass-dependent length scale $L_{CS}$ associated with the system. {\it Unlike what is done in the standard treatment}, we here construct the following action
\begin{equation}
\frac{S}{\hbar} = \frac{1}{L_{Pl}^2} \int d^4x \;\sqrt{-g} R + \frac{L_{CS}^2}{L_{Pl}^2}\int d^{4}x \; \sqrt{-g}\;\overline{\psi}\; i\gamma^{\mu}\partial_\mu\psi 
- \frac{m_{Pl}c}{\hbar} \frac{L_{CS}}{2L_{Pl}}\int d^{4}x\; \sqrt{-g}\; \overline{\psi}{\psi}
\label{actualaction}
\end{equation}
If the second term can be neglected compared to the first one, and we set $L_{CS}/L_{Pl}=2m_c/m_{Pl}\gg 1$
then the above action reduces to
\begin{equation}
S_E = \frac{c^3}{G} \int d^4x \;\sqrt{-g} R - m_c c \int d^{4}x\; \sqrt{-g}\; \overline{\psi}{\psi}
\label{Eisteinaction}
\end{equation}
from which Einstein equations can be obtained after replacing $\bar{\psi}\psi$  with the  delta function
$\delta^{3}({\bf x})$. If the first term can be neglected compared to the second one, and we set $L_{CS}/L_{Pl}=m_{Pl}/2m_{q}\gg 1$, then the above action reduces to
\begin{equation}
 \frac{4m_q^2}{m_{Pl}^2}\; {S_D} =\hbar \int d^{4}x   \sqrt{-g}\;\overline{\psi}\; i\gamma^{\mu}\partial_\mu\psi 
- {m_{q}c} \int d^{4}x\; \sqrt{-g}\; \overline{\psi}{\psi}
\label{Diracaction}
\end{equation}
Although the action gets scaled by a constant, its variation nonetheless yields the Dirac equation.

In this sense the Dirac equation for a mass $m_q \ll m_{Pl}$ and the Einstein equations for a mass $m_c\gg m_{Pl}$ are dual to each other, and involve the same length $L_{CS}$. They both arise from the same underlying action (\ref{actualaction}) provided
$L_{CS}/L_{Pl} = 2m_c/m_{Pl}=m_{Pl}/2m_q$, i.e.   $m_q m_c=m_{Pl}^2/4$. This is true not only for the special form (\ref{unilength}) for $L_{CS}$ but for any generic form which limits to $\lambda_C$ and $R_S$ at the two ends, and has a minimum around $m_{Pl}$. Variation of the action (\ref{actualaction}) with respect to the metric yields generalised Einstein equations, and variation with respect to the Dirac state yields a modified Dirac equation which now depends on Planck length. The properties of these equations are worth investigating, from the point of view of singularity resolution and other issues. A mass $m$ now `knows', unlike before, whether to obey the Dirac equation or the Einstein equation or an entirely new equation: this is determined by the action (\ref{actualaction}).  The action (\ref{actualaction}) can be written more compactly and elegantly as
\begin{equation}
\frac{L_{Pl}^2 }{\hbar} S =  \int d^4x \;\sqrt{-g} \left[ R\;  - \frac{1}{2}\; L_{CS}\; \overline{\psi}{\psi} \; + \; L_{CS}^2 \; \overline{\psi}\; i\gamma^{\mu}\partial_\mu\psi 
\right] 
\label{actualaction2}
\end{equation}
Much in the spirit of the AdS/CFT correspondence, one can hope to learn about macroscopic black holes by studying their dual Dirac particles, because both have the same length $L_{CS}$, and hence obey the `same' physics, according to the action (\ref{actualaction}). 

One can verify from the form of the action (\ref{actualaction}) that the first term is indeed negligible compared to the second one, for $m\ll m_{Pl}$. The integrand of the first term behaves as $R_S/r^{3}$, whereas the
integrand of the second term behaves as $\lambda_C /r^3$ [the length scale $L_{CS}\sim \lambda_C$, which is of the same order as the scale of the gradient, and we assume $\overline{\psi}\psi\sim 1/r^3$]. The required condition is hence satisfied since $R_S\ll \lambda_C$. On the other hand, when $m\gg m_{Pl}$, the second term vanishes on scales $R_S$ and larger, since the state $\psi$ is strongly confined on the scale $\lambda_C$, and $\lambda_C\ll R_S$. 

One could now proceed to study the equations that follow from (\ref{actualaction}), but that perhaps is not the complete story. While $L_{CS}$ acts as a binding agent between quantum theory and gravity, the two fields, the dominating Dirac field on the left of the curve in Fig.1, and gravity on the right, are very different mathematical entities. One being a four-component spinor and another a second rank tensor field: how does one expect a transition from one to the other as one crosses $m_{Pl}$? A common mathematical language is highly desirable, and is in fact provided by the Newman-Penrose formalism, where the Riemann tensor is expressed in terms of the so-called Ricci rotation coefficients, using the so-called Ricci identities \cite{Chandra}. When this is done, the Einstein equations begin to look remarkably similar to Dirac equations written in the same formalism. Motivated by this similarity, we made the radical suggestion that the Dirac spinors be identified with Ricci rotation coefficients. Dirac equations can then be written in a manner similar to Einstein equations, with the Dirac mass acting as a source for the Ricci coefficients \cite{Singh2014a}.

This however comes at a price. The Dirac equations land up satisfying severely undesirable constraints. Remarkably enough, the constraints all disappear entirely if one introduces torsion in the space-time, and identifies the Dirac field with the {\it complex} torsion  part of the rotation coefficients. This is independent support for torsion, which supplements the geometric motivation we gave above. We thus have gravity, described by the torsion free part of the Ricci coefficients, and the Dirac field, which is described by the torsion part. This seems like a nice way to bring together gravity and quantum theory, but while this has been done \cite{Singh2014a}, earlier on we did not have the newly discovered length scale $L_{CS}$ and the related action principle (\ref{actualaction2}). In forthcoming work we will cast this new action principle in the Newman-Penrose formalism, and investigate closely the consequences of this duality between torsion and gravity.

Black holes mysteriously appear similar to elementary particles, both possessing the same set of conserved charges: mass, electric charge, and angular momentum. In fact it is known that all black hole solutions belong to Petrov Class D \cite{Chandra}, whereas the Dirac particles are in a sense duals of this Petrov type \cite{Singh2014a}. However, up until now, this similarity/duality  only seemed like a coincidence. Through the newly discovered length scale $L_{CS}$ reported here, and through the common underlying action principle, we expect to arrive at a better understanding of how and why such a similarity arises in the first place. 

\newpage

\centerline{\bf REFERENCES}

\bigskip
\setstretch{1.2}


\end{document}